\documentclass[twocolumn,showpacs,preprintnumbers,amsmath,amssymb,floatfix,prc]{revtex4}
\usepackage{graphicx}
\usepackage{dcolumn}
\usepackage{bm}

\def\beq{\begin{equation}}
\def\eeq{\end{equation}}
\def\be{\begin{eqnarray}}
\def\ee{\end{eqnarray}}

\newcommand{\lsim}{
 \mathrel{\setbox0=\hbox{$<$}\raise0.6ex\copy0\kern-\wd0
 \lower0.65ex\hbox{$\sim$}}}

\newcommand{\gsim}{
 \mathrel{\setbox0=\hbox{$>$}\raise0.6ex\copy0\kern-\wd0
 \lower0.65ex\hbox{$\sim$}}}

\begin{document}
\title{Constraining Gluon Shadowing Using Photoproduction in 
Ultraperipheral pA and AA Collisions}
\author{Adeola Adeluyi}
\author{C.A. Bertulani}
\affiliation{Department of Physics \& Astronomy,
Texas A\&M University-Commerce, Commerce, TX 75428, USA}

\date{\today}
\begin{abstract}
Photoproduction of heavy quarks and exclusive production of vector
mesons in ultraperipheral proton-nucleus and nucleus-nucleus
collisions depend significantly on nuclear gluon distributions. 
In the present study we investigate quantitatively the extent of the 
applicability of these processes at the Large Hadron Collider (LHC) in 
constraining the shadowing component of nuclear gluon modifications. 
\end{abstract}
\maketitle
\vspace{1cm}
%
%
\section{Introduction}
Ultraperipheral relativistic heavy ion collisions can explore several
aspects of particle and nuclear physics and have been extensively
discussed in the literature (for a small sample of references, see
e.g. \cite{Jackson, Bertulani:1988,Cahn:1990jk,Baur:1990fx,KN99,
Bertulani:1999cq,Goncalves:2001vs,KNV02,Goncalves:2003is,Bertulani:2005ru,
Baltz:2007kq,AyalaFilho:2008zr,Salgado:2011wc}). 
In a recent publication we explored lead-lead peripheral collisions 
at energies available at the CERN Large Hadron Collider to probe gluon 
distributions in nuclei \cite{Adeluyi:2011rt}. Using recent gluon 
distributions from global fits to data, we investigated the 
sensitivity of direct photoproduction of heavy quarks and exclusive 
production of vector mesons to varying strength of gluon
modifications. This idea, originally proposed in
Ref. \cite{Goncalves:2001vs}, can be used to constrain nuclear gluon
distribution from data on production of heavy quarks and vector mesons.
 
In this article we extend our previous work by considering both
direct and resolved processes in the photoproduction of heavy quarks 
in ultraperipheral proton-lead (pPb) and lead-lead (PbPb) collisions 
at the LHC. We also consider, as an addition to the previous treatment, 
exclusive production of vector mesons in pPb collisions. Our main goal 
is to investigate quantitatively the extent of applicability of these 
processes in constraining nuclear gluon modifications. In order to 
present a self-contained report we include all relevant results and 
discussions from our previous study.

The paper is organized as follows: below we briefly discuss the photon 
flux generated by one of the nuclei (or proton), making reference to
several previous publications where it has been discussed in details.
Sec.~\ref{phothadupc} treats the photoprocesses considered in this
study, as well as the input nuclear and photon parton distributions.  
In Sec.~\ref{res} we present the results of our calculations and a 
brief comment on theoretical errors. Our conclusion is contained in 
Sec.~\ref{conc}.  

For a given impact 
parameter ${\bf b}$, the flux of virtual photons with photon energy
$k$, ${d^3N_\gamma(k,{\bf b}) / dkd^2b} $,  is strongly dependent
on the Lorentz factor $\gamma$. At the Large Hadron Collider (LHC) at 
CERN the Lorentz factor in the laboratory frame $\gamma_L$ is $7455$ 
for proton-proton (pp), $4690$ for proton-lead (pPb) and $2930$ 
for lead-lead (PbPb) collisions. The relationship between the Lorentz 
contraction factor associated with the relative velocity between the 
colliding nuclei, and the collider energy per nucleon, $E/A$, in GeV, 
is given by $\gamma=2\gamma_L^2-1 \approx 2(1.0735E/A)^2$.
The photon flux also depends strongly on the  adiabaticity parameter 
 $\zeta=kb/\gamma$ \cite{Bertulani:1988, Cahn:1990jk,Baur:1990fx}:
\begin{equation}
\frac{d^3N_\gamma(k,{\bf b})}{dkd^2b} = 
\frac{Z^2\alpha \zeta^2}{\pi^2kb^2} \left[ K_1^2(\zeta) + \frac{1}{
\gamma_L^2} K_0^2(\zeta) \right] \, \, ,
\label{dpf}
\end{equation}
which drops off exponentially for $\zeta >1$, above a cutoff 
energy determined  essentially by the size of the nucleus, 
$E_{cutoff} \sim \gamma$MeV$/b$ (fm).

For symmetric nucleus-nucleus (AA) collisions, such as PbPb
collisions at the LHC, each nucleus can act equally as source or 
target of the photon flux. 
Integrating ${d^3N_\gamma(k,{\bf b}) / dkd^2b} $ over impact
parameters with the constraint of no hadronic interactions  and 
accounting for the photon polarization yields the total 
photon flux ${dN_\gamma^Z(k) / dk} $  given by  \cite{KN99,Baltz:2007kq},
\begin{eqnarray}
\frac{dN_\gamma^Z(k)}{dk}& =&  
2 \pi \int_{2R_A}^{\infty} db \, b 
\int_0^R  {dr \, r \over \pi R_A^2} 
\int_0^{2\pi} d\phi  \nonumber \\
&\times& {d^3N_\gamma(k,b+r\cos \phi)\over dkd^2b} \, \, ,
\label{npf} 
\end{eqnarray}  
with $R_A$ the radius of the nucleus.
  
In the case of proton-nucleus (pA) collisions, the nucleus 
acts preferentially as the source and the proton as the target, 
leading predominantly to $\gamma$p processes. But there is still 
a non-negligible contribution from $\gamma$A processes in which the 
proton acts as the source of photons and the nucleus as the target. 
Thus expressions for both types of fluxes are required for pA collisions.
The flux due to the nucleus (of charge Z) can be evaluated 
analytically and is given by \cite{Bertulani:1988},
\begin{eqnarray}
\frac{dN_\gamma^Z(k)}{dk} &=& 
{2Z^2 \alpha \over\pi k} \bigg[ \zeta_R^{pA}K_0(\zeta_R^{pA})
K_1(\zeta_R^{pA}) \nonumber \\
&-& {(\zeta_R^{pA})^2\over 2} \big(K_1^2(\zeta_R^{pA})-K_0^2(\zeta_R^{pA})
\big) \bigg] \, \, , 
\label{anaf}
\end{eqnarray}
with reduced adiabaticity parameter, $\zeta_R^{pA}$, given by
$\zeta_R^{pA} = k(R_p + R_A)/\gamma$ and $R_p$ the effective radius 
of the proton. 

The flux due to the proton is usually estimated using the dipole
formula for the electric form factor \cite{Drees:1988pp}: 
\begin{eqnarray}
\frac{dN_\gamma^p(k)}{dk} &=& \frac{\alpha}{2\pi k} 
\bigg[1+ \big(1-\frac{2k}{\sqrt{S_{NN}}}\big)^2\bigg]
\nonumber \\
&&\bigg(\ln{D} - \frac{11}{6} + \frac{3}{D} - \frac{3}{2D^2}
+\frac{1}{3D^3}\bigg) \, \, ,
\label{pf}
\end{eqnarray}
where
$D = 1 + [\rm 0.71 \, GeV^2/Q_{\rm min}^2]$ and the minimum momentum 
transferred $Q_{\rm min}^2 = k^2/[\gamma^2(1-2k/\sqrt{S_{NN}})]$.

With the knowledge of the photon flux, any generic total photoproduction 
cross section can be expressed as a convolution
of a process cross section $\sigma_{\gamma}^{X}(k)$ 
and the photon flux, $dN_\gamma/dk$. Thus for AA collisions 
\begin{equation}
\sigma^{X}=2 \int dk \,\frac{dN_\gamma^Z(k)}{dk}\, \sigma^{\gamma A
  \rightarrow X}(k) \, \, ,
\label{tAAxs}
\end{equation}
with ${dN_\gamma^Z}/{dk}$ given by  Eq.~(\ref{npf}). The factor of $2$ 
accounts for the source/target symmetry present in AA collisions. In
the case of pA collisions, we have
\begin{equation}
\sigma^{X}= \int dk \bigg[\frac{dN_\gamma^Z}{dk}\, \sigma^{\gamma p
  \rightarrow X}(k)
+ \frac{dN_\gamma^p}{dk}\, \sigma^{\gamma A
  \rightarrow X}(k)\bigg] \, \, ,
\label{tpAxs}
\end{equation}
with ${dN_\gamma^Z}/{dk}$ and ${dN_\gamma^p}/{dk}$ given by  
Eq.~(\ref{anaf}) and Eq.~(\ref{pf}) respectively. Relevant 
expressions for the rapidity distributions for both AA and pA
collisions are given later.
\section{Photon-hadron interactions in ultraperipheral collisions}
\label{phothadupc}
\subsection{Photoproduction of heavy quarks}
From the viewpoint of a Fock space decomposition, photon
interactions with hadrons and nuclei can be classified as 
direct or resolved.
In direct interactions the photon behave as a point-like
particle (``bare photon'') while in resolved interactions the 
photon fluctuates into a quark-antiquark state or an even more complex 
partonic configuration consisting of quarks and gluons. The cross section 
for the photoproduction of a pair of heavy quarks, 
$\sigma^{\gamma H\to Q\overline{Q}X}\, (k)$,
is thus a sum of both the direct and resolved 
contributions,
\begin{equation}
\sigma^{\gamma H \rightarrow Q\overline{Q}X}\, (k) = 
\sigma^{\gamma H \rightarrow Q\overline{Q}X}_{direct}\, (k) + 
\sigma^{\gamma H \rightarrow Q\overline{Q}X}_{resolved}\, (k) .
\end{equation}
Here $H$ stands for a proton or a nucleus ($H \equiv p, A$)
and the total photoproduction cross section is obtained by convoluting 
the equivalent photon flux, $dN_\gamma(k)/dk$, with 
$\sigma^{\gamma H\rightarrow Q\overline{Q}X}\, (k)$, as described by 
Eq.~(\ref{tAAxs}) and Eq.~(\ref{tpAxs}) for AA and pA collisions.

Let us now consider both processes in some detail.
At leading order (LO), photon-gluon fusion leading to the production of 
a heavy quark pair is the only subprocess relevant to direct 
photoproduction. In view of the high energies
involved, perturbative QCD is applicable, and the direct 
photoproduction cross section 
can be expressed as a convolution of the partonic cross section for the
subprocess $\gamma g \rightarrow Q \overline{Q}$ and the relevant
nucleon or nuclear gluon distribution:
\begin{equation}
\sigma^{\gamma H\rightarrow Q\overline{Q}X}_{direct}\, (s) =
\int dx
 \,  \sigma^{\gamma g \rightarrow Q\overline{Q}}(\hat{s}) \,
 f_{g}^{H}(x,Q^2)\,\Theta(\zeta) \, ,
\label{dppcs}
\end{equation}
with $x$ the momentum fraction carried by the gluon.
We use $m_Q$ for the mass of the heavy
quark (charm or bottom), $s=W_{\gamma H}^2$ denotes 
the square of the center-of-mass energy of the photon-nucleus 
(or photon-nucleon) system, $\hat{s}=W_{\gamma g}^2$ that of the 
photon-gluon system, and $\zeta = \hat{s}-4m_{Q}^2$. 
The gluon distribution in $H$, $f_{g}^{H}(x, Q^2)$, is evaluated at 
the pQCD factorization scale $Q^2=W_{\gamma g}^2=\hat{s}$. 
The function $\Theta(\zeta)$ enforces a minimum (``threshold'') value 
of $x$, $x_{min}$, on the integral given by 
$x_{min} = 4m_q^2/W_{\gamma H}^2$ 

The partonic photon-gluon fusion cross section is given 
by \cite{Gluck:1978bf,JonesWyld,FriStreng78}
\begin{eqnarray}
\sigma^{\gamma g \rightarrow Q \overline{Q}} \, (\hat{s})= \frac {2\pi\,\alpha_{em}\,\alpha_s(Q^2)\,e_Q^2}
{\hat{s}} \nonumber \\ 
\times \left[(1+\beta -\frac{\beta^2}{2}) 
\ln\big[\frac{1+\nu}
{1-\nu}\big]  -(1+\beta)\, \nu\right] \,\,,
\label{plcs}
\end{eqnarray}
with $e_Q$ the electric charge of the heavy quark $Q$, $\alpha_{em}$ the 
electromagnetic coupling constant, 
$\beta=4\,m_Q^2/\hat{s}$, and $\nu=\sqrt{1-\beta}$.

The resolved contribution is identical to the hadroproduction of 
heavy quarks, and at leading order, involves terms corresponding to 
gluon-gluon and quark-antiquark subprocesses. The resolved cross
section can be written as
\begin{widetext}
\begin{equation}
\sigma^{\gamma H\rightarrow Q\overline{Q}X}_{resolved}\, (s) =
\int d\tilde{x} \left[f_{g}^{\gamma}(x_1,Q^2) f_{g}^{H}(x_2,Q^2)
 \sigma^{g g \rightarrow Q\overline{Q}}(\hat{s}) 
+ \sum_{q}f_{q}^{\gamma}(x_1,Q^2)[f_{q}^{H}(x_2,Q^2)+f_{\bar{q}}^{H}(x_2,Q^2)]
\sigma^{q \bar{q} \rightarrow Q\overline{Q}}(\hat{s}) \right]\Theta(\zeta)
\label{rppcs}
\end{equation}
\end{widetext}
where $d\tilde{x} \equiv dx_1 dx_2$, $f_{a}^{\gamma}(x_1,Q^2)$ ($f_{a}^{H}(x_2,Q^2)$)
is the distribution of parton $a$ with momentum fraction $x_1$ 
($x_2$) in a photon ($H$) respectively, $\hat{s}=x_1x_2s$
and $\zeta = \hat{s}-4m_{Q}^2$. The summation over $q$ runs over the
light flavors, i.e. $q = u,d,s$.

The partonic cross sections $\sigma^{g g \rightarrow  Q\overline{Q}}(\hat{s})$  
and $\sigma^{q \bar{q} \rightarrow Q\overline{Q}}(\hat{s})$ are given
by \cite{Gluck:1977zm,Combridge:1978kx,Brock:1993sz}
\begin{eqnarray}
&& \sigma^{g g \rightarrow Q \overline{Q}} \, (\hat{s})= \frac {\pi \,\alpha_s^2(Q^2)}
{3\hat{s}} \nonumber \\ 
&\times&\left[(1+\beta +\frac{\beta^2}{16}) \ln(\frac{1+\nu}{1-\nu})  
-(\frac{7}{4}+\frac{31}{16}\beta)\, \nu\right] \,
\label{ggcs}
\end{eqnarray}
and
\begin{eqnarray}
\sigma^{q \bar{q} \rightarrow Q \overline{Q}} \, (\hat{s})= \frac {8\pi\,\alpha_s^2(Q^2)}
{27 \hat{s}}\left[(1+\frac{\beta}{2}) \sqrt{1-\beta}\right] \,\, .
\label{qqbcs}
\end{eqnarray}
Here $\beta$ and $\nu$ are as defined previously.

It is often more enlightening to represent the 
cross section in terms of rapidity, that is, to consider
rapidity distributions.
The differential cross section with respect to the rapidity 
of the heavy quark pair, 
$d\sigma/dy$, is related to the differential cross section
with respect to photon energy, $d\sigma/dk$, through 
the relation $d\sigma/dy = kd\sigma/dk$. The rapidity 
distribution can therefore be expressed as 
\begin{equation}
\frac{d \sigma^{\gamma H \rightarrow Q \overline{Q}X}}{dy} = 
k\, \frac{dN_\gamma(k)}{dk}\, \sigma^{\gamma H
\rightarrow Q \overline{Q}X}(k)
\label{Hqqhrap}
\end{equation}
and scales directly with the photon flux $dN_\gamma/dk$.
Thus for pA collisions, with the convention that the proton is
incident from the right and the nucleus from the left, the total 
rapidity distribution is
\begin{eqnarray}
\frac{d \sigma^{pA \rightarrow Q \overline{Q}X}}{dy}&=& 
\bigg[k \, \frac{dN_\gamma^Z(k)}{dk}\, \sigma^{\gamma p
\rightarrow Q \overline{Q}X}(k)\bigg]_{k=k_l} \nonumber \\
&+&\bigg[k \, \frac{dN_\gamma^p(k)}{dk}\, \sigma^{\gamma A
\rightarrow Q \overline{Q}X}(k)\bigg]_{k=k_r} 
\label{pAqqhrap}
\end{eqnarray}
where $k_l$ ($k_l \propto e^{-y}$) and $k_r$ ($k_r \propto e^{y}$) 
simply denote photons from the nucleus and proton respectively,
and the fluxes ${dN_\gamma^Z}/{dk}$ and ${dN_\gamma^p}/{dk}$ are given by  
Eq.~(\ref{anaf}) and Eq.~(\ref{pf}). As remarked earlier, the 
fluxes have support only at small values of $k$, dying out 
exponentially at large $k$. Thus the first term on the right-hand side
($\gamma$p distribution) peaks at positive rapidities while the 
second term ($\gamma$A distribution) peaks at negative rapidities.
Since both the fluxes and process cross sections are different,
the total distribution is manifestly asymmetric, and the $\gamma$p 
term dominates due to the much larger nuclear 
flux ${dN_\gamma^Z}/{dk}$. 

The total rapidity distribution for AA collisions can 
likewise be written as 
\begin{eqnarray}
\frac{d \sigma^{AA \rightarrow Q \overline{Q}X}}{dy}&=& 
\bigg[k \, \frac{dN_\gamma^Z(k)}{dk}\, \sigma^{\gamma p
\rightarrow Q \overline{Q}X}(k)\bigg]_{k=k_l} \nonumber \\
&+&\bigg[k \, \frac{dN_\gamma^Z(k)}{dk}\, \sigma^{\gamma A
\rightarrow Q \overline{Q}X}(k)\bigg]_{k=k_r} 
\label{AAqqhrap}
\end{eqnarray}
with $k_l$ ($k_r$) simply denoting photons from the nucleus 
incident from the left (right) and the flux ${dN_\gamma^Z}/{dk}$
given by Eq.~(\ref{npf}). Here the process cross sections and
the left/right fluxes are identical, thus the respective rapidity 
distributions are mirror images of each other, and consequently the 
total distribution is symmetric about midrapidity ($y=0$). 

\subsection{Exclusive production of vector mesons}
The cross section for the exclusive elastic photoproduction of a
vector meson $V$ on $H$ ($H \equiv p, A$) can be written as
\begin{equation}
\sigma^{\gamma H \rightarrow VH} (k) = \left.  \frac{d \sigma^{\gamma
H \rightarrow VH}}{dt} \right| _{t=0}  \int dt | F_H(t) |^{2} ,
\label{eepvm}
\end{equation}
where $d \sigma^{\gamma A \rightarrow VA}/ dt|_{t=0}$ is the forward
scattering amplitude and
$F_H(t)$ is the form factor. The dynamical
information is encoded in the forward scattering amplitude while
the momentum transfer of the elastic scattering is determined by the
form factor, which is, in general, dependent on the spatial attributes 
of the target $H$. 

There have been studies of the photoproduction of $J/\Psi$ and  
$\Upsilon$ in ultraperipheral collisions at LHC (see for instance \cite{KN99,
Goncalves:2001vs,Baltz:2007kq,
strikman_plb,per4,ivanov_kop,vicmag_prd2008,AyalaFilho:2008zr,
strikman_jhep,klein_prl,vicmag_prc}).
In this work we use the simple amplitude calculated from leading order 
two-gluon exchange in perturbative QCD \cite{Ryskin,Brodsky:1994kf}
and corrected for other 
relevant effects (such as relativistic corrections, inclusion of the
real part of the scattering amplitude, next-to-leading order NLO
effects, etc, see for instance \cite{Ryskin:1995hz,Frankfurt:1997fj}) through a 
phenomenological multiplicative correction factor $\zeta_V$. 

For elastic photoproduction on protons, the corrected LO 
scattering amplitude can be written as
\begin{equation}
\left.  \frac{d \sigma^{\gamma p \rightarrow Vp}}{dt} \right| _{t=0} =
\zeta_V \frac{16\pi^3 \alpha_{s} ^{2} \, \Gamma_{ee}}{3 \alpha M_{V} ^{5}}
\left[  x g_p(x,Q^{2}) \right] ^{2} .
\label{epvm_p}
\end{equation}
Here, $M_V$ is the mass of the vector meson ($J/\Psi$ and
$\Upsilon(1s)$ in the present study), $x = M_V^2/W_{\gamma p}^2$ is the fraction of the nucleon momentum carried by the
gluons, and $g_p(x,Q^{2})$ is the gluon distribution in a proton,
evaluated at a momentum transfer $Q^{2} = (M_{V}/2)^{2}$. 
$\Gamma_{ee}$ is the leptonic decay width and $\alpha_s$ 
($\alpha$) the strong (electromagnetic) coupling constant.
Eq.~\ref{epvm_p} is easily generalized to the nuclear case,
\begin{equation}
\left.  \frac{d \sigma^{\gamma A \rightarrow VA}}{dt} \right| _{t=0} =
\zeta_V \frac{16\pi^3 \alpha_{s} ^{2} \, \Gamma_{ee}}{3 \alpha M_{V} ^{5}}
\left[  x g_A(x,Q^{2}) \right] ^{2} ,
\label{epvm_A}
\end{equation}
where $g_A(x,Q^{2}) = g_p(x,Q^{2}) \times R_g^A(x,Q^{2})$ is the
nuclear gluon distribution and $ R_g^A(x,Q^{2})$ the gluon modification.

The correction factor $\zeta_V$ is estimated by constraining the 
calculated cross sections for elastic vector meson photoproduction on protons, 
$\sigma^{\gamma p \rightarrow Vp}(W_{\gamma p})$, to reasonably
reproduce the photoproduction data from HERA: \cite{Adloff:2000vm} 
for $J/\Psi$ and
\cite{Adloff:2000vm,Breitweg:1998ki,Chekanov:2009zz} for $\Upsilon
(1s)$. Further details can be found in \cite{Adeluyi:2011rt}.
In contrast to photoproduction of heavy quarks, the quadratic 
dependence of the cross section on the gluon distribution
has the significant implication of making exclusive vector
meson production a very sensitive probe of nuclear gluon
modifications. 

We now discuss the momentum-squared transferred dependence ($t$-dependence) 
of the cross section. For a proton the standard practice 
is to parametrize the $t$-dependence in the form of a rapidly
decreasing exponential function, $e^{b|t|}$, where $b$ is referred to
as the slope parameter. Integrating this function with respect to 
$t$ gives a multiplicative factor $1/b$ with units GeV$^{2}$. 
In the case of a nucleus, the nuclear form factor $F_A(t)$ is given 
by the Fourier transform of the nuclear density distribution:
$F_A(t) = \int d^3r \, \rho_A(r) \, e^{i{\bf q} \cdot {\bf r}}$,
where $q$ is the momentum transferred. 
For a heavy nucleus it is customary to model the density
distribution as a Woods-Saxon distribution with parameters from 
electron scattering, $\rho_A(r)= \rho_0 /[1 + e^{[(r-R_{A})/d]}]$,
with central density $\rho_0$, radius $R_{A}$ and skin depth $d$.
For $^{208}$Pb in use at the LHC,
$\rho_0 = 0.16$/fm$^{3}$, $R_{A} = 1.2 A^{1/3}$ fm, and $d =
0.549$ fm \cite{DeJager:1974dg}. 

For the proton the photonucleon cross section
$\sigma^{\gamma p \rightarrow Vp} (k)$ is obtained through 
\begin{equation}
\sigma^{\gamma p \rightarrow Vp} (k) = \frac{1}{b}\, \frac{d\sigma^{\gamma p\rightarrow
Vp}}{dt}\bigg|_{t=0}
\end{equation}
with slope parameter $b$. In the present study we employ $b = 4.5$ GeV$^{-2}$.
The photonuclear cross section is given by
\begin{equation}
\sigma^{\gamma A\rightarrow VA}(k) = \frac{d\sigma^{\gamma A\rightarrow
VA}}{dt}\bigg|_{t=0} \int_{t_{min}(k)}^\infty dt |F(t)|^2
\end{equation}
Here $t_{min}(k)=(M_v^2/4k\gamma_{L})^2$, as is appropriate for narrow 
resonances \cite{klein_prl}.
The total cross sections for pA and AA collisions are obtained 
by a convolution of
$\sigma^{\gamma p \rightarrow Vp} (k)$ and $\sigma^{\gamma  A\rightarrow VA}(k)$ 
with the relevant photon flux as described by 
Eq.~(\ref{tpAxs}) and Eq.~(\ref{tAAxs}) respectively.

The photon energy, $k$, is related to the rapidity of the vector
meson $y$ by $k= (M_{V}/2) \exp(y)$. 
Thus, as in the case of photoproduction of heavy quarks, 
the differential cross section with respect to rapidity is given by
${d \sigma^{\gamma H \rightarrow VH}}/{dy} =
( k {dN_\gamma(k)}/{dk}) \sigma^{\gamma H \rightarrow VH}(k)$, and 
the corresponding expressions for rapidity distributions 
in pA and AA collisions are of the forms given by 
Eq.~(\ref{pAqqhrap}) and Eq.~(\ref{AAqqhrap}) respectively. 
The symmetry attributes of the distributions are also similar.

\subsection{Parton distributions in nuclei and photons}
We now turn to the consideration of the parton distributions (PDs) relevant to 
the processes considered in the present study. As mentioned earlier, 
the direct contribution to the photoproduction of heavy quarks is 
dependent solely on the gluon distributions in protons and nuclei,
same as for the elastic photoproduction of vector mesons. The
resolved contribution, on the other hand, requires the distributions
of light quarks and antiquarks and gluons in photons, protons and 
nuclei. Thus in addition to the usual requirement of nucleon and 
nuclear parton distributions (nPDs), there is also the need for the 
relatively poorly known parton distributions in photons ($\gamma$PDs), thereby 
increasing the level of the theoretical uncertainties in the
calculation of photoproduction of heavy quarks. Although the direct 
contribution is dominant, this dominance does not vitiate the need to 
have a good control on the resolved contribution, especially for 
$b\bar{b}$ production where the resolved components are quite sizable.

Let us first discuss nuclear parton distributions. It is rather well-known
that the distributions of partons (i.e. quarks and gluons) in 
nuclei are quite different from the distributions in free nucleons,
that is, they are ``modified'' by the complex, many-body effects in the 
nucleus. These nuclear effects are usually 
parametrized in terms of "nuclear modifications" $R_{a}^A(x,Q^2)$ 
which in general depend on the parton specie ($a$), the nucleus ($A$), 
momentum fraction $x$ and scale $Q^2$.  
The nuclear effects can be categorized based on different intervals in $x$.
At small values of $x$ ($x \lesssim 0.04$), we have the phenomenon
generally referred to as shadowing. This is a depletion, in the sense
that in this interval, the distribution of a parton $a$ in the nucleus
is smaller compared to the corresponding distribution in a free 
proton, i.e. $R_a^A < 1$. Antishadowing, which is an
enhancement ($R_a^A > 1$), occurs in the range $0.04 \lesssim x \lesssim
0.3$. Another depletion, the classic EMC effect \cite{Aubert:1983xm}, is
present in the interval $0.3 \lesssim x \lesssim 0.8$, while for 
$x > 0.8$, the Fermi motion region, we have another enhancement. 
It is important to note that although both shadowing and
the EMC effect (antishadowing and Fermi motion) correspond to
depletion (enhancement), the physical principles and mechanisms 
governing these phenomena are quite different. Further details can be
found in \cite{Geesaman:1995yd,Piller:1999wx,Armesto:2006ph,Kolhinen:2005az} 
With the knowledge of $R_{a}^A(x,Q^2)$, nuclear parton distributions 
can be expressed as a convolution of free nucleon parton distributions 
and nuclear modifications, i.e. 
$f_a^A(x,Q^2) = f_a(x,Q^2) \otimes R_{a}^A(x,Q^2)$.

While the determination of quark and antiquark distributions in 
nucleons and nuclei is in general a nontrivial task, that of gluons
is even more problematic. Gluons are electrically neutral, and 
thus their distributions cannot be extracted directly from 
Deeply Inelastic Scattering (DIS) and Drell-Yan (DY) processes 
which account for the major part of the data used in global fits.  
Their distributions are in general inferred from sum
rules and the $Q^2$ evolution of sea quarks distributions. 
The situation is even worse in the nuclear case: the
available data is much less than for nucleons, and there is the added
complication of a mass dependence. It is therefore not unusual for
nuclear gluon distributions from different global fits to differ 
significantly, especially in the magnitude of the various nuclear
effects (shadowing, antishadowing, etc). This is especially obvious
at low $Q^2$ (i.e. around their initial starting scales) since 
evolution to high $Q^2$ tends to lessen the differences.  
Earlier global analyses 
\cite{Eskola:1998df,deFlorian:2003qf,Shad_HKN,Hirai:2007sx} relied
heavily on fixed-target nuclear deep-inelastic scattering
(DIS) and Drell-Yan (DY) lepton-pair production. 
Incorporation of data on inclusive hadron production in deuteron-gold 
collisions has been implemented in \cite{Eskola:2008ca,Eskola:2009uj},
and neutrino-iron data in \cite{Schienbein:2009kk,Stavreva:2010mw,
Kovarik:2010uv}. We should also mention 
the approach in \cite{Frankfurt:2003zd,Frankfurt:2011cs} utilizing the Gribov 
picture of shadowing. Despite all these advances the nuclear gluon
distribution is still currently the least constrained aspect of global fits to 
nuclear parton distributions, as significant uncertainties still persist at 
both small and large $x$.
    
Four recent nucleon and nuclear parton distributions are utilized in 
the present study. For the proton we use the Martin-Stirling-Thorne-Watts
(MSTW08) parton distributions \cite{Martin:2009iq} which are 
available up to next-to-next-to-leading order (NNLO).
In the nuclear case we use three nuclear modification sets.
Two sets are by  Eskola, Paukunnen, and Salgado, namely EPS08 and EPS09 
\cite{Eskola:2008ca,Eskola:2009uj}. 
The third is the Hirai-Kumano-Nagai (HKN07) distributions 
\cite{Hirai:2007sx}. While EPS08 is only to leading order (LO), 
both EPS09 and HKN07 are available up to next-to-leading order (NLO). 
The distributions from MSTW08 serve two purposes: 
as the free nucleon distributions used in conjunction with
nuclear modifications, and also as a ``special'' nuclear 
distribution in the absence of nuclear effects. The latter case is
particularly useful for highlighting the influence of the various
nuclear effects on observables. 

\begin{figure}[!htb]
\begin{center} 
\includegraphics[width=8.75cm, height=8.5cm, angle=270]{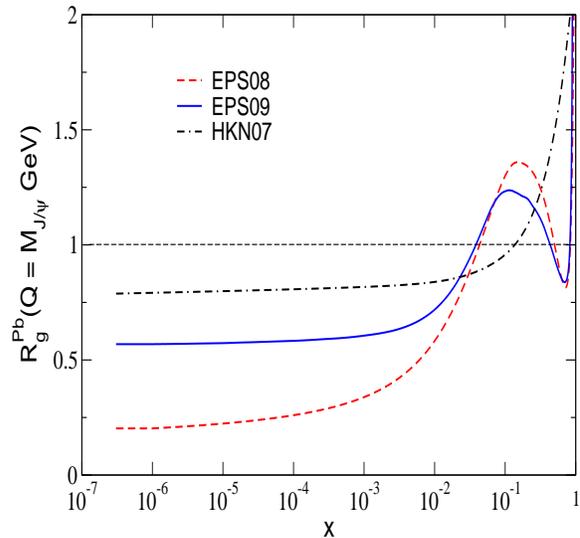}
\end{center}
\caption[...]{(Color online) Nuclear gluon modifications in Pb,
 $R_{g}^{Pb}(x,Q^2 = M_{J/\Psi}^2)$, from EPS08 (dashed line), 
 EPS09 (solid line), and HKN07 (dash-dotted line) respectively.}
\label{fig:RgPb_Mjpsi}
\end{figure}

It is instructive to compare the characteristics of the gluon
distributions from the four aforementioned sets based on the strength 
of their nuclear modifications. In Fig.~\ref{fig:RgPb_Mjpsi} we show 
the nuclear modifications for gluons in Pb, $R_g^Pb(x,Q^2)$ from 
EPS08, EPS09, and HKN07 at the factorization scale $Q^2 =
M_{J/\Psi}^2$, appropriate for the elastic photoproduction of the $J/\Psi$ 
meson. As already stated, one can view the MSTW08 gluon distribution 
as a nuclear gluon distribution in the limit of zero nuclear effects 
( $R_a^A = 1$).
At this scale, HKN07 has a rather weak gluon shadowing which extends
well into the antishadowing region, no antishadowing and  
gluon EMC effect, and an early onset of Fermi motion. 
EPS09 exhibits a moderately strong shadowing, and appreciable
antishadowing and EMC effect, with a quite strong Fermi motion. 
Nuclear modifications are strongest in EPS08: an especially strong 
shadowing, and substantial antishadowing, EMC, and Fermi motion. 
Thus in terms of shadowing we have 
a progression from zero effects to weak effects, moderate
(intermediate) effects, then to strong effects as one progresses from 
MSTW08 to EPS08. The issue of uncertainties in parton distributions 
is becoming increasingly important especially in relation to
precision tests of QCD. Discussions on uncertainties in
nuclear parton distributions can be found in \cite{Hirai:2007sx,Eskola:2009uj}. 

Parton distributions in photons ($\gamma$PDs) are derived from experimentally 
determined photon structure function $F_2^{\gamma}(x,Q^2)$, in conjunction 
with appreciable theoretical inputs. These inputs, which are 
necessary in implementing the parametrizations of photon parton 
distributions from the structure function, account in part for 
some of the observable differences in the available photon 
parton distribution sets. Another source of differences is in the 
choice and scope of experimental data from which $F_2^{\gamma}$ is 
extracted. At present there is an appreciable number of photon parton 
distribution sets available, both at leading and next-to-leading orders
\cite{Duke:1982bj,Drees:1984cx,Abramowicz:1991yb,Hagiwara:1994ag,
Gluck:1991ee,Gluck:1994tv,Gluck:1991jc,Gordon:1996pm,Gordon:1991tk,
Schuler:1996fc,Schuler:1995fk,Aurenche:1994in,Aurenche:1992sb,Cornet:2002iy,
Cornet:2004ng,Cornet:2004nb}. It should be noted that  
unlike in the case of a nucleon, there are no valence quarks 
present in the photon; therefore antiquark distributions are the 
same as quark distributions. Furthermore, there are no sum rules 
governing the gluon content; thus the gluon distribution is 
almost totally unconstrained. The gluon distribution contributes to 
$F_2^{\gamma}(x,Q^2)$ majorly through the $\gamma^*g \to q\bar{q}$ 
channel, which has significant numerical support only at small $x$.
 
\begin{figure}[!htb]
\includegraphics[width=8.75cm, height=9.25cm, angle=270]{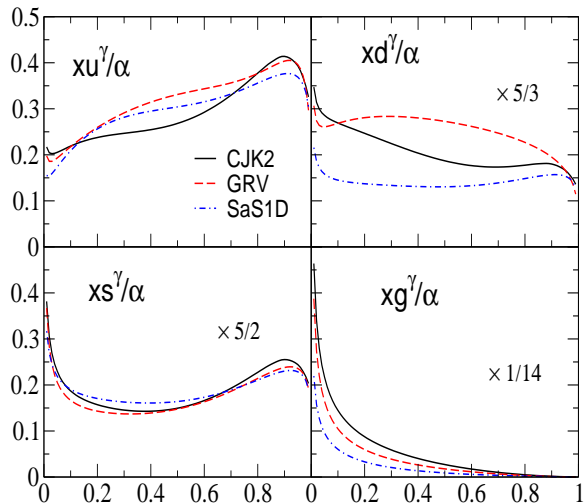}
\caption[...]{(Color online) Parton distributions in the photon 
for light quarks (u,d,s) and gluons at $Q^2 = 10$ GeV$^2$ 
for three distribution sets:  CJK2 (solid line) \cite{Cornet:2004ng}, 
GRV (dashed line) \cite{Gluck:1991jc}, and SaS1d (dash-dotted line) 
\cite{Schuler:1995fk} respectively. For better visuality the down 
quark distributions have been enhanced by a factor of $5/3$, the strange quark
distributions by $5/2$, and the gluon distributions by $1/14$.}
\label{photpdf3d}
\end{figure}

In light of the relatively poor situation especially in the gluon
sector, we use three different photon parton distribution sets in the
present study. The rationale for this will become clear when we
present results for the resolved component in heavy quark
production. Two sets are relatively older and have been in common 
use: the Gluck-Reya-Vogt (GRV) set \cite{Gluck:1991jc} and that 
by Schuler and Sjostrand (SaS1d) \cite{Schuler:1995fk}. 
The third set, the Cornet-Jankowski-Krawczyck (CJK2)
set \cite{Cornet:2004ng}, is more recent and has facility to estimate 
error on a calculated quantity due to the uncertainties in the input photon
parton distributions. In Fig.~\ref{photpdf3d} we show the parton 
distributions from these three sets for the light (u,d,s) quarks and gluon.  

The up quark is the best constrained component of photon parton 
distributions, and even in this case there are clear differences 
between the three sets. This is also the case for the down quark.
Only for the strange quark distribution is there good agreement 
between the three sets. But these differences pale in comparison 
with that of gluons where the differences between the sets are quite
significant almost across the whole $x$ interval, with CJK2 having 
the largest magnitude while SaS1d has the lowest, with the GRV 
somewhat intermediate. For the LO resolved contribution to heavy 
quark photoproduction, the two competing partonic subprocesses are 
the $gg \to Q\bar{Q}$ and $q\bar{q} \to Q\bar{Q}$, and it thus to be 
expected that significant differences will arise in the resolved cross 
sections whenever the $gg \to Q\bar{Q}$ component is appreciable.

\section{Results}
\label{res}
We now discuss the results of our calculations for both the inclusive 
photoproduction of heavy quarks ($c\bar{c}$ and $b\bar{b}$) and the 
exclusive elastic production of vector mesons ($J/\Psi$ and 
$\Upsilon(1s)$) in pPb ($\sqrt{s_{NN}}=8.8$ TeV) and PbPb 
($\sqrt{s_{NN}}=5.5$ TeV) collisions at the LHC. 
We take $m_c = 1.4$ GeV and $m_b = 4.75$ GeV for
consistency with the MSTW08 parton distributions, and 
$m_{J/\Psi} = 3.097$ GeV and $m_{\Upsilon} = 9.46$ GeV respectively. 
The strong coupling constant $\alpha_{s}(Q^{2})$, needed for the 
calculations, is evaluated to one loop at the scale $Q^{2}$
using the evolution code contained in the MSTW08 package.
Two comments pertaining to the results on photoproduction of heavy quarks 
are in order: first, due to the relative smallness of nuclear effects 
in heavy quark photoproduction, especially for pPb collisions, we 
have, for visual purposes, left out the results from the HKN07 
distributions. In general the HKN07 values are intermediate between 
those of EPS09 and MSTW08. Second, while we have used three sets of 
photon parton distributions in the calculations, we show only the 
rapidity distribution plots using the GRV set, since the GRV results
are consistently somewhat intermediate between those from CJK2 and 
SaS1d distributions.

\subsection{Heavy Quarks in pA and AA Collisions}
\subsubsection{Open charm}
In Table~\ref{cccbarpA} we present the cross sections for both 
direct and resolved $c\bar{c}$ production in ultraperipheral 
pPb collisions at the LHC, using the previously-described
nucleon/nuclear and photon parton distributions. Due to the 
large disparity in the magnitude of the photon fluxes from the 
proton and the nucleus (Pb), it is a common practice to neglect 
the photonuclear ($\gamma$Pb) contribution to the total cross
section. Another simplification is the neglect of the resolved 
contribution which is relatively small in comparison with the 
dominant direct contribution. Since both the $\gamma$Pb and resolved 
contributions are dependent on nuclear and photon parton
distributions, the results presented in 
Table~\ref{cccbarpA} afford a good handle in determining the 
degree of validity of the above approximations for the rather 
significantly different attributes of the distributions utilized 
in the present study. 
\begin{table}[!htb]
\caption{Cross sections for photoproduction of 
$c\bar{c}$ in ultraperipheral pPb collisions at the LHC. All
cross sections are in microbarns ($\mu$b).}
\begin{tabular}[c]{|l| l c c| c c| c c c c|}
\hline
       &&PDF            && Direct       &&&& Resolved & \\ 
\cline{7-10}
        &&                &&             && SaS1d   && GRV      & CJK   \\
\hline
$\gamma$p &&MSTW08          && 5570      && 692   && 1157   & 1418  \\
\hline
         &&MSTW08          && 607       && 114   && 195    & 228   \\
$\gamma$A &&EPS08           && 376       && 94    && 162    & 185  \\
         &&EPS09	    && 471       && 102   && 175    & 202  \\
\hline
\end{tabular} 
\label{cccbarpA}
\end{table}

We first consider the photonuclear part. The $\gamma$Pb
contribution to the total cross section is approximately $10.6\%$ for 
MSTW08 (no nuclear modifications), $8.8\%$ for EPS09 (
moderate gluon shadowing), and $7.4\%$ for EPS08 (strong 
gluon shadowing), almost totally independent of the choice of 
photon parton distribution set. The resolved contribution, on 
the other hand, while depending strongly on the choice of 
photon parton distributions, has an even more negligible dependence 
on the choice of nuclear parton distributions. For the SaS1d
distribution, the resolved component is approximately $12\%$ 
while it is approximately $18\%$ for the GRV distribution, and 
$21\%$ for the CJK2 set. As is to be expected about $96\%$ of the
resolved component stems from $\gamma$p while about $90\%$ of the 
$\gamma$Pb contribution derives from the direct component.
Overall the sensitivity to nuclear modifications (dominantly shadowing) is 
rather small: the no-modification (MSTW08) total cross section is
reduced by approximately $4\%$ and $2\%$ by the modifications in 
EPS08 and EPS09 respectively. 

\begin{figure}[!htb]
\begin{center} 
\includegraphics[width=8.5cm, height=8.5cm, angle=270]{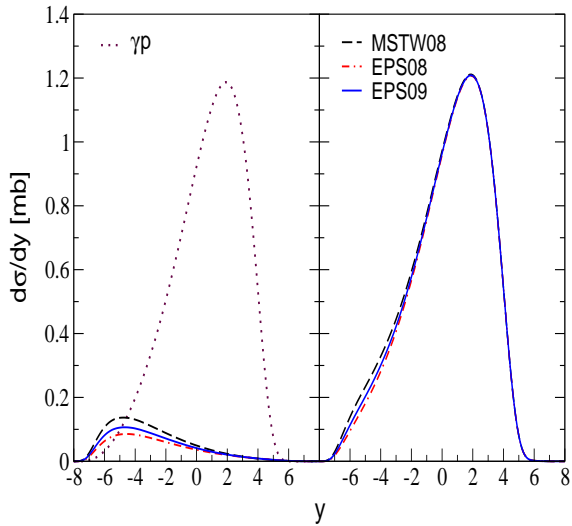}
\end{center}
\caption[...]{(Color online) Rapidity distributions of $c\bar{c}$ 
photoproduction in pPb collisions at the LHC using the GRV photon 
parton distributions. Left panel: $\gamma$p and $\gamma$Pb
contributions to total rapidity distributions. Dotted line depicts 
the $\gamma$p contribution while the dashed (MSTW08), solid 
(EPS09), and dash-dotted (EPS08) lines correspond to $\gamma$Pb 
contributions with no shadowing, moderate shadowing, and strong
shadowing respectively. Right panel: total rapidity distributions 
(sum of $\gamma$p and $\gamma$Pb contributions).}
\label{fig:pphqpArapcc}
\end{figure}

In Fig.~\ref{fig:pphqpArapcc} we show the corresponding rapidity 
distributions for $c\bar{c}$ production using the GRV photon parton 
distributions. 
The left panel depicts the distributions separately for both $\gamma$p 
and $\gamma$Pb contributions while the right panel shows the total 
distributions, i.e. the sum of both contributions. As is apparent 
from the left panel, the $\gamma$p contribution is dominant at all 
rapidities except for $y \lesssim -5$. In line with our convention,
the $\gamma$p contribution peaks at positive rapidities while 
the $\gamma$Pb distributions peak at negative rapidities, and the 
asymmetric nature of the total distributions is clearly exhibited. 
Despite the smallness of the $\gamma$Pb contribution, the effects of nuclear 
modifications are still slightly manifested in the total rapidity 
distributions shown in the right panel. It is interesting to note that 
the sensitivity to nuclear effects is significantly enhanced when 
one considers the ratio of the total rapidity distribution at a 
specific negative rapidity ($-y_1$), and the rapidity distribution
at the equivalent positive rapidity ($+y_1$), somewhat reminiscent 
of pseudorapidity asymmetry in deuteron-gold collisions. The MSTW08 
ratio for $y=-3,3$ is reduced by about $9\%$ and $5\%$ by the 
modifications in EPS08 and EPS09 respectively, while for $y=-4,4$ 
the reductions amount to $15\%$ and $10\%$, and $23\%$ and $14\%$ 
respectively for $y=-5,5$. Thus the rapidity asymmetry ratio in the 
interval $-5 \le y \le 4, 4 \le y \le 5$ exhibits appreciable 
sensitivity to nuclear effects.

We now discuss the corresponding case of $c\bar{c}$ production in 
ultraperipheral PbPb collisions, and in Table~\ref{cccbarAA} we show the 
cross sections (in mb) for both direct and resolved contributions.
\begin{table}[!htb]
\caption{Cross sections for photoproduction of 
$c\bar{c}$ in ultraperipheral PbPb collisions at the LHC. All
cross sections are in millibarns (mb).}
\begin{tabular}[c]{|l|c c c| c c c c c c |}
\hline
 PDF            && Direct              &&&&& Resolved && \\ 
\cline{5-10}
                &&             &&& SaS1d   && GRV      && CJK   \\
\hline
MSTW08          && 1167     &&& 110   && 180    && 226  \\
EPS08           && 890       &&& 99    && 165    && 207  \\
EPS09	        && 1002      &&& 101   && 169    && 213  \\
\hline
\end{tabular} 
\label{cccbarAA}
\end{table}
 
The sensitivity to nuclear modifications (dominantly shadowing) is
more significant here than in pPb collisions. Thus considering 
the direct contribution which is the dominant part, the 
modification-free (MSTW08) cross section is reduced by approximately 
$24\%$ by the rather strong nuclear shadowing present in EPS08, and by 
about $14\%$ in the case of EPS09. With the inclusion of the resolved 
component, the reductions are around $22\%$ and $13\%$ respectively,
and vary only slightly with the choice of photon parton distribution.

The resolved contribution exhibits a mild dependence on nuclear 
modifications and a much stronger sensitivity to photon parton
distribution. Thus in the case of SaS1d the resolved component 
is $8.6\%$ of the total cross section for MSTW08, $10\%$ for 
EPS08, and $9.2\%$ for EPS09. For GRV, the contributions are 
$13.4\%$, $15.6\%$, and $14.4\%$ respectively. The CJK2 values 
are the largest: $16.2\%$, $18.9\%$, and $17.5\%$ respectively. 

\begin{figure}[!htb]
\begin{center} 
\includegraphics[width=8.5cm, height=8.5cm, angle=270]{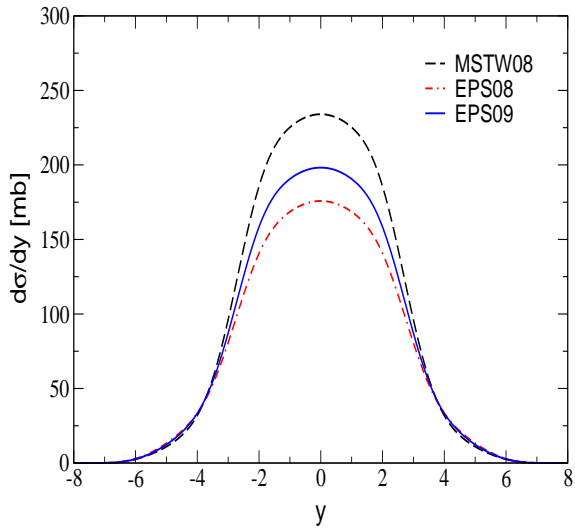}
\end{center}
\caption[...]{(Color online) Total rapidity distributions of the
  photoproduction of $c\bar{c}$ in PbPb collisions at the LHC using 
  the GRV photon parton distributions. Dashed
  line depicts result using the MSTW08 gluon distribution (no nuclear
  modifications). Solid and dash-dotted lines are results
  from nuclear-modified gluon distributions from EPS09 and EPS08
  respectively.}
\label{thqrapcc}
\end{figure}

The rapidity distributions for $c\bar{c}$ production in
ultraperipheral PbPb collisions are shown in Fig.~\ref{thqrapcc}, 
and are manifestly symmetric about midrapidity ($y = 0$).
The sensitivity to nuclear modifications is more transparent here than 
in total cross sections. 
%
Shadowing is the dominant nuclear effect for $-3 < y < 3$, and the rapidity
distributions in this region reproduce the observed trend of 
gluon shadowing strength exhibited in  Fig.~\ref{fig:RgPb_Mjpsi}.
MSTW08 with its zero gluon shadowing gives the largest rapidity
distribution while EPS08, with its strong gluon shadowing, gives the
smallest. Due to strong flux suppression, shadowing is most markedly 
apparent for the rapidity range $-2 \lesssim y \lesssim 2$. 
This range therefore provides a good window to discriminate among 
different gluon shadowing scenarios.

The rapidity intervals  $3 < y < 6$ 
corresponds to $x_{min}$ in the antishadowing region (deep shadowing) 
for right (left) incident photons and vice versa for $-6 < y < -3$. 
Due to the photon flux suppression 
in the deep shadowing region, the rapidity distributions are 
sensitive mainly to  antishadowing in addition to both EMC effect and 
Fermi motion. Since both EPS08 and EPS09 have substantial
antishadowing, their rapidity distributions reflect this, 
being slightly larger than those from MSTW08. The
discriminatory power here is not as appreciable as in the shadowing
case though, due largely to the smallness of the distributions.

For both rapidity ranges $y < -6$ and $y > 6$, $x_{min} > 0.2$ and the 
relevant contributing nuclear effects are the EMC and Fermi motion
since the contribution from antishadowing is small, and that from
shadowing practically nonexistent by virtue of flux suppression. 
Both EPS08 and EPS09 nuclear
modifications exhibit EMC effect and Fermi motion, and the destructive 
interference from both effects render their rapidity distributions to
practically coincide with that from MSTW08.   
\subsubsection{Open bottom}
We now discuss our results for total cross sections and 
rapidity distributions for $b\bar{b}$ production.  
In Table~\ref{cbbbarpA} we present the cross sections for both 
direct and resolved $b\bar{b}$ production in ultraperipheral 
pPb collisions at the LHC. In parallel with the treatment of 
$c\bar{c}$ production, we determine the relative importance 
of $\gamma$Pb and resolved contributions to the total $b\bar{b}$
production cross sections.
\begin{table}[!h]
\caption{Cross sections for photoproduction of 
$b\bar{b}$ in ultraperipheral pPb collisions at the LHC. All
cross sections are in nanobarns (nb).}
\begin{tabular}[c]{|l| l c c| c c| c c c c|}
\hline
       &&PDF            && Direct       &&&& Resolved & \\ 
\cline{7-10}
        &&                &&             && SaS1d   && GRV      & CJK   \\
\hline
$\gamma$p &&MSTW08        && 36512      && 8641   && 12178   & 14977  \\
\hline
         &&MSTW08          && 5084       && 2061   && 3032    & 3663  \\
$\gamma$A &&EPS08          && 3972       && 1886    && 2808    & 3389  \\
         &&EPS09	   && 4409      && 1936   && 2874   & 3477  \\
\hline
\end{tabular} 
\label{cbbbarpA}
\end{table}
For MSTW08 (no nuclear modifications) the $\gamma$Pb
contribution to the total cross section is approximately $14\%$, 
$13\%$ for EPS09 ( moderate gluon shadowing), and $12\%$ for 
EPS08 (strong gluon shadowing), with little dependence on the choice of 
photon parton distribution set. It is thus apparent that while the
photonuclear contribution is relatively more significant here than in 
$c\bar{c}$ production the effect of nuclear shadowing is less. 
The resolved contribution again depends significantly 
on the choice of photon parton distribution, and exhibits negligible dependence 
on the choice of nuclear parton distribution. For the SaS1d
distribution, the resolved component is approximately $21\%$ 
while it is approximately $27\%$ for the GRV distribution, and 
$30\%$ for the CJK2 set. These values are larger than the 
corresponding ones for $c\bar{c}$ production, and thus serve to 
underscore the relatively higher importance of the resolved component 
in $b\bar{b}$ photoproduction. 
As noted above, the sensitivity to nuclear modifications 
(dominantly shadowing) is very small: the no-modification (MSTW08) 
total cross section is reduced by approximately $2.4\%$ and $1.5\%$ 
by the modifications in EPS08 and EPS09 respectively. Thus with its 
appreciably large resolved component and weak sensitivity to nuclear 
effects, it is tempting to speculate that $b\bar{b}$ photoproduction 
could be of some use in constraining photon parton distributions.
\begin{figure}[!htb]
\begin{center} 
\includegraphics[width=8.5cm, height=8.5cm, angle=270]{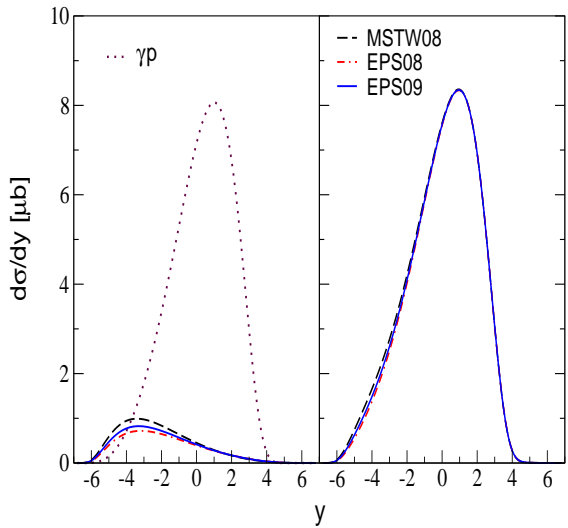}
\end{center}
\caption[...]{(Color online) Rapidity distributions of $b\bar{b}$ 
photoproduction in pPb collisions at the LHC using the GRV photon 
parton distributions. Left panel: $\gamma$p and $\gamma$Pb
contributions to total rapidity distributions. Dotted line depicts 
the $\gamma$p contribution while the dashed (MSTW08), solid 
(EPS09), and dash-dotted (EPS09) lines correspond to $\gamma$Pb 
contributions with no shadowing, moderate shadowing, and strong
shadowing respectively. Right panel: total rapidity distributions 
(sum of $\gamma$p and $\gamma$Pb contributions).}
\label{fig:pphqpArapbb}
\end{figure}         

In Fig.~\ref{fig:pphqpArapbb} we show the corresponding rapidity 
distributions for $b\bar{b}$ production using the GRV photon parton 
distributions. 
The left panel depicts the distributions separately for both $\gamma$p 
and $\gamma$Pb contributions while the right panel shows the total 
distributions, being the sum of both contributions. As is apparent 
from the left panel, the $\gamma$p contribution is dominant at all 
rapidities except for $y \lesssim -4$. As in $c\bar{c}$ production
the $\gamma$p contribution peaks at positive rapidities while 
the $\gamma$Pb distributions peak at negative rapidities, and the 
asymmetric nature of the total distributions is clearly manifested. 
Due to the smallness of the nuclear effects, the total rapidity 
distributions shown in the right panel almost overlap for the 
different nuclear parton distributions considered in the present 
study. Similar to $c\bar{c}$ production, the rapidity asymmetry 
ratio exhibits appreciable sensitivity to nuclear effects around 
$y=-4,4$: the MSTW08 ratio is reduced by $19\%$ and $6\%$ respectively 
by the modifications in EPS08 and EPS09.

Let us now consider the corresponding case of $b\bar{b}$ production in 
ultraperipheral PbPb collisions. In Table~\ref{cbbbarAA} we show the 
cross sections (in $\mu$b) for both direct and resolved contributions.  
\begin{table}[!htb]
\caption{Cross sections for photoproduction of 
$b\bar{b}$ in ultraperipheral PbPb collisions at the LHC. All
cross sections are in microbarns ($\mu$b).}
\begin{tabular}[c]{|l|c c c| c c c c c c |}
\hline
 PDF            && Direct              &&&&& Resolved && \\ 
\cline{5-10}
                &&             &&& SaS1d   && GRV      && CJK   \\
\hline
MSTW08          && 6227      &&& 1076   && 1468    && 1800  \\
EPS08           && 5812       &&& 982    && 1378    && 1734  \\
EPS09	        && 5992      &&& 969   && 1357    && 1708  \\
\hline
\end{tabular} 
\label{cbbbarAA}
\end{table}

The sensitivity to nuclear modifications, while more appreciable here 
than in pPb collisions, is significantly less than for the equivalent 
$c\bar{c}$ production. 
The modification-free (MSTW08) cross section is reduced  
$7\%$ by the modifications in EPS08, and  
about $5\%$ in the case of EPS09. 
The resolved contribution is almost independent of nuclear 
modifications but exhibits a strong sensitivity to photon parton
distribution. Thus the resolved component 
is approximately $15\%$ of the total cross section for SaS1d, $19\%$ 
for GRV, and about $23\%$ for CJK2. Again, these values indicate that 
the resolved component is relatively more significant in $b\bar{b}$
as compared to $c\bar{c}$ photoproduction.  

\begin{figure}[!htb]
\begin{center} 
\includegraphics[width=8.5cm, height=8.5cm, angle=270]{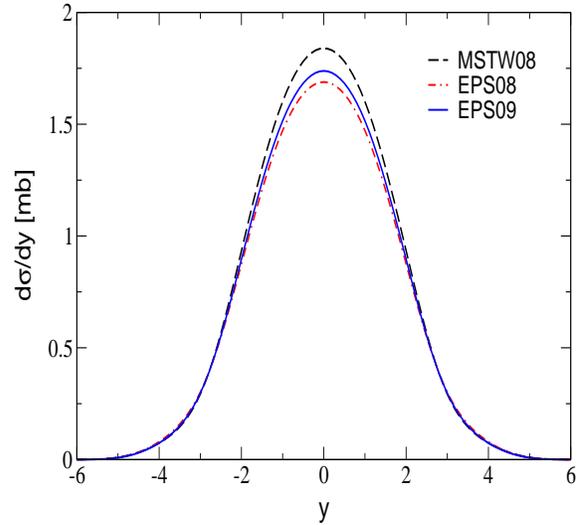}
\end{center}
\caption[...]{(Color online) Total rapidity distributions of $b\bar{b}$ 
  photoproduction in PbPb collisions at the LHC using the GRV photon
  parton distributions. Dashed
  line depicts result using the MSTW08 gluon distribution (no nuclear
  modifications). Solid and dash-dotted lines are results
  from nuclear-modified distributions from EPS09 and EPS08 respectively.}
\label{fig:hqrapbb}
\end{figure}

In Fig.~\ref{fig:hqrapbb} we show the rapidity distribution for $b\bar{b}$
in ultraperipheral PbPb collisions at the LHC. 
Shadowing dominates in the rapidity interval $-2 < y < 2$ and is 
most clearly manifested in the rapidity window $-1 < y < 1$. Thus 
this interval presents the best sensitivity to shadowing effects 
in $b\bar{b}$ production. Although less marked, the progression 
of rapidity distribution with relative shadowing strength follows the 
trend observed in $c\bar{c}$ production: MSTW08 still gives the
largest distribution while EPS08 gives the smallest.
As in the case of $c\bar{c}$ production, there is a slight
manifestation of the influence of antishadowing around  
$y = -3$ and $y = 3$. The distributions practically overlap
for $y < -4$ and $y > 4$; thus overall, the interval 
$-1 \lesssim y \lesssim 1$ seems to afford the best sensitivity to 
nuclear effects, in this case primarily shadowing. A detailed 
treatment of $b\bar{b}$ production at the LHC with emphasis on 
the $x$ dependence can be found in \cite{Strikman:2005yv}. 
\subsection{Vector Mesons in pA and AA Collisions} 
\subsubsection{$J/\Psi$}
We now present our results on elastic photoproduction of the $J/\Psi$ 
and $\Upsilon(1s)$ in the framework of a corrected leading-order 
two-gluon exchange formalism in QCD. Here the nuclear effect
(shadowing) is quite sizable; we therefore also display explicitly results 
from the HKN07 distribution set.

In Table~\ref{tjpsipPb} we present the components and total cross 
sections for the elastic photoproduction of  $J/\Psi$ in
ultraperipheral pPb collisions at the LHC. As expected the $\gamma$p 
contribution is dominant, with the $\gamma$Pb contribution decreasing 
in size with increasing severity of gluon shadowing. 
\begin{table}[!htb]
\caption{Cross sections (in $\mu$b) for elastic photoproduction of 
 $J/\Psi$ in ultraperipheral pPb collisions at the LHC. Second 
and third columns are the contributions from $\gamma$p and 
$\gamma$Pb respectively for different distributions. The sums of the
two contributions are presented in the fourth column.}
\begin{tabular}[c]{|l|ccc c c c c c c c| c c c c}
\hline
    &             && $\gamma$p    &&& $\gamma$A    &&& Total & \\
\hline
MSTW08&           && 167.3        &&& 23.6         &&& 190.9 & \\
EPS08&            &&              &&& 2.2          &&& 169.5 & \\		       
EPS09&	         &&              &&& 8.5          &&& 175.8  & \\		      
HKN07&            &&              &&& 15.4         &&& 182.7  & \\	
\hline               
\end{tabular}
\label{tjpsipPb}
\end{table}

The contribution of the $\gamma$Pb component to the total cross
section ranges from a high of about $12\%$ for MSTW08 (no gluon shadowing) 
to a low of $1.3\%$ for EPS08 (strong gluon shadowing), in
accordance with the trend displayed in Fig.~\ref{fig:RgPb_Mjpsi}. This 
trend is replicated in the degree of shadowing reflected in the total
cross section: the no-shadowing (MSTW08) cross section is reduced by 
about $11\%$ by the shadowing in EPS08, $8\%$ in EPS09, and $4\%$ in 
HKN07 respectively.  

\begin{figure}[!htb]
\begin{center} 
\includegraphics[width=8.5cm, height=8.5cm, angle=270]{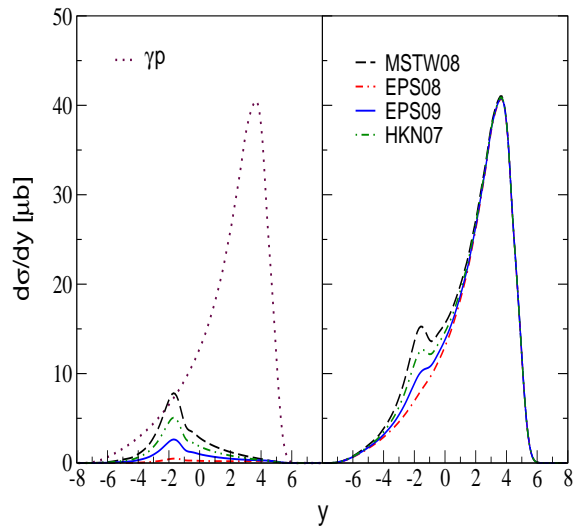}
\end{center}
\caption[...]{(Color online) Rapidity distributions of exclusive
photoproduction of $J/\Psi$ in pPb collisions at the LHC. 
Left panel: $\gamma$p and $\gamma$Pb
contributions to total rapidity distributions. Dotted line depicts 
the $\gamma$p contribution while the dashed (MSTW08), dash-double-
dotted (HKN07), solid 
(EPS09), and dash-dotted (EPS08) lines correspond to $\gamma$Pb 
contributions with no shadowing, weak shadowing, moderate shadowing, 
and strong shadowing respectively. Right panel: total rapidity distributions 
(sum of $\gamma$p and $\gamma$Pb contributions).}
\label{jpsirappPb}
\end{figure}

Fig.~\ref{jpsirappPb} shows the rapidity distributions for both 
$\gamma$p and $\gamma$Pb components and their sums. The distributions 
are manifestly asymmetric, with the dominant $\gamma$p component 
peaking at positive rapidities while the subdominant $\gamma$Pb 
contributions peak at negative rapidities.
The $\gamma$Pb distributions exhibit clearly the influence of 
gluon shadowing, especially in the narrow rapidity window 
$-3 \lesssim y \lesssim 1$ where the differences mimic the 
relative strength of gluon shadowing in the respective nuclear 
parton distribution. Although the $\gamma$p contribution 
is dominant for practically all rapidities, these $\gamma$Pb 
distributions with the exception of EPS08 are appreciable enough 
in the said rapidity range such that the total rapidity distributions 
show the same tendency as that of the $\gamma$Pb component in the 
designated rapidity window.
Thus it seems feasible that a consideration of $J/\Psi$ production 
in pPb collisions in this rapidity interval offers some potential 
in constraining gluon shadowing. 

We now turn to $J/\Psi$ production in utraperipheral PbPb collisions
at the LHC and in Table~\ref{tjpsiPbPb} we present the total cross
sections from the different parton distributions. The differences in 
the predicted cross sections are extremely clear cut, and thus this 
process can serve as an excellent probe of gluon shadowing as well as 
a good discriminator of the different gluon shadowing templates
considered in the current study.
\begin{table}[!htb]
\caption{Total cross sections (in mb) for elastic photoproduction of 
 $J/\Psi$ in ultraperipheral PbPb collisions at the LHC.}
\begin{tabular}[c]{|lc|c|c c c}
\hline
PDF                 && Cross Section \\
\hline
MSTW08              && 74 \\		      
EPS08               && 10 \\		       
EPS09	            && 29 \\		      
HKN07               && 49 \\
\hline	               
\end{tabular}
\label{tjpsiPbPb}
\end{table}

It is instructive to consider the remarkable effect that the 
quadratic gluon dependence has on the cross 
sections (and also on the rapidity distributions). Thus the 
no shadowing cross section is reduced by approximately $87\%$ 
by the strong shadowing in EPS08, and by about $61\%$ by the 
relatively moderate shadowing in EPS09. Even the rather weak 
gluon shadowing present in HKN07 is responsible for reducing the 
cross section by about $34\%$. 

\begin{figure}[!htb]
\begin{center} 
\includegraphics[width=8.5cm, height=8.5cm, angle=270]{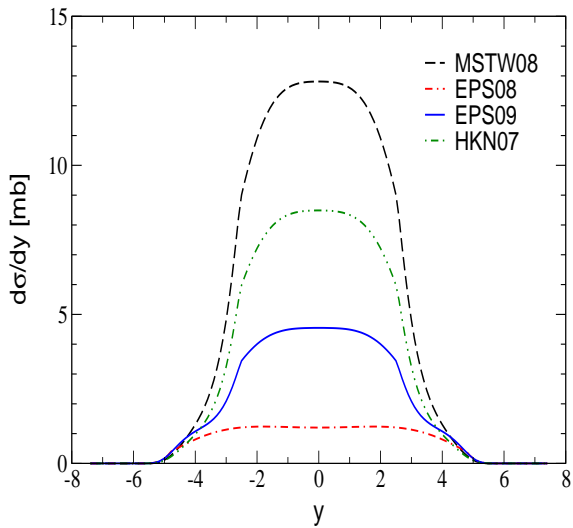}
\end{center}
\caption[...]{(Color online) Total rapidity distributions of exclusive
  photoproduction of $J/\Psi$ in PbPb collisions at the LHC. Dashed
  line depicts result using the MSTW08 gluon distribution (no nuclear
  modifications). Solid, dash-dotted, and dash-double-dotted lines are results
  from nuclear-modified gluon distributions from EPS09, EPS08, and
  HKN07 parton distributions respectively.}
\label{jpsirapPbPb}
\end{figure}

The constraining potential of the $J/\Psi$ elastic production 
process is more clearly exhibited when rapidity distributions are 
considered. To this end we display the corresponding rapidity 
distributions in  Fig.~\ref{jpsirapPbPb} for the four different 
gluon shadowing scenarios. 

Shadowing is the relevant nuclear effect in the rapidity interval 
$-3 < y < 3$ and unsurprisingly, the rapidity distributions mimic 
the behavior in the shadowing region of Fig.~\ref{fig:RgPb_Mjpsi}. 
The largest rapidity distribution is given by MSTW08, followed by
HKN07, and EPS09. The smallest is by EPS08 due to its strong gluon 
shadowing. The rapidity window $-2 < y < 2$ manifestly depicts the 
significant distinction between the various gluon distributions 
arising from the quadratic dependence. Antishadowing manifests 
in the intervals $-5 < y < -4$ and $4 < y < 5$; the effect though 
is quite slight.      

\subsubsection{$\Upsilon(1s)$}
We now discuss our results for elastic production of {$\Upsilon(1s)$ 
in ultraperipheral pPb and PbPb collisions. The ensuing treatment 
parallels closely that of $J/\Psi$ in many respects due to the same 
underlying production mechanism.

In Table~\ref{tupsipPb} we present the components and total cross 
sections for the elastic photoproduction of  $\Upsilon$ in
ultraperipheral pPb collisions at the LHC. While the $\gamma$p 
contribution is still dominant, the $\gamma$Pb contributions
from the different gluon distributions are markedly significant,
and exhibit the trend of decreasing magnitude with increasing 
strength of gluon shadowing. 

\begin{table}[!htb]
\caption{Total cross sections (in nb) for elastic photoproduction of 
 $\Upsilon$ in ultraperipheral pPb collisions at the LHC. Second 
and third columns are the contributions from $\gamma$p and 
$\gamma$Pb respectively for different distributions. The sums of the
two contributions are presented in the fourth column.}
\begin{tabular}[c]{|l|ccc c c c c c c c| c c c c}
\hline
    &             && $\gamma$p    &&& $\gamma$A    &&& Total & \\
\hline
MSTW08&           && 390        &&& 219         &&& 609 & \\
EPS08&            &&              &&& 84          &&& 474 & \\		       
EPS09&	         &&              &&& 130          &&& 520  & \\		      
HKN07&            &&              &&& 161         &&& 551  & \\	
\hline               
\end{tabular}
\label{tupsipPb}
\end{table}

The contribution of the $\gamma$Pb component to the total cross
section is $36\%$ for MSTW08 (no gluon shadowing), $29\%$ for HKN07 
(weak gluon shadowing), $25\%$ for EPS09 (moderate gluon shadowing), 
and $18\%$ for EPS08 (strong gluon shadowing). This gradation accords
with the trend observed in Fig.~\ref{fig:RgPb_Mjpsi}, which is also  
replicated in the degree of shadowing reflected in the total
cross section: the no-shadowing (MSTW08) cross section is reduced by 
about $22\%$ by the shadowing in EPS08, $15\%$ in EPS09, and $10\%$ in 
HKN07 respectively. This appreciable magnitude of the effect of shadowing
seems to indicate that $\Upsilon$ photoproduction cross section
in ultraperipheral pPb collisions offers some promising potential in
constraining nuclear gluon shadowing.    

In Fig.~\ref{upsirappPb} we show the rapidity distributions for both 
$\gamma$p and $\gamma$Pb components (left panel )and the total 
(right panel). As usual the distributions are asymmetric, with the 
dominant $\gamma$p component peaking at positive rapidities while the 
subdominant $\gamma$Pb contributions peak at negative rapidities
according to the convention adopted in the present study.
\begin{figure}[!htb]
\begin{center} 
\includegraphics[width=8.5cm, height=8.5cm, angle=270]{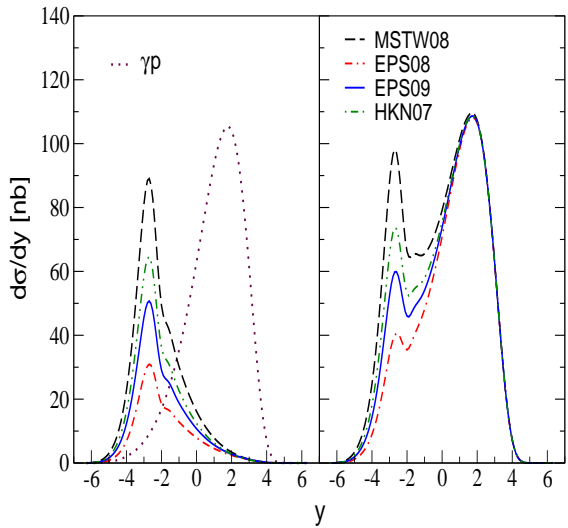}
\end{center}
\caption[...]{(Color online) Rapidity distributions of exclusive
photoproduction of $\Upsilon(1s)$ in pPb collisions at the LHC.
Left panel: $\gamma$p and $\gamma$Pb
contributions to total rapidity distributions. Dotted line depicts 
the $\gamma$p contribution while the dashed (MSTW08), dash-double-
dotted (HKN07), solid 
(EPS09), and dash-dotted (EPS08) lines correspond to $\gamma$Pb 
contributions with no shadowing, weak shadowing, moderate shadowing, 
and strong shadowing respectively. Right panel: total rapidity distributions 
(sum of $\gamma$p and $\gamma$Pb contributions).}
\label{upsirappPb}
\end{figure}

Let us consider the left panel. The $\gamma$Pb component distributions here, 
relative to that of $\gamma$p, are more appreciable than for $J\Psi$ 
production. In fact, for $-5 \lesssim y \lesssim -2$ these
distributions are larger than the $\gamma$p distribution, and around
their peaks at $y \sim -2.7$ they reflect quite distinctly the effect
of the varying shadowing strength in the gluon distributions used.    
Thus in the rapidity interval $-4 \lesssim y \lesssim -1$ the 
total distributions in the right panel show good sensitivity to 
gluon shadowing, and therefore afford good potential for constraining 
purposes.

We now turn to $\Upsilon$ production in ultraperipheral PbPb collisions
at the LHC. In Table~\ref{tupsiPbPb} we present the total cross
sections from the different gluon distributions. 
\begin{table}[!htb]
\caption{Total cross sections (in $\mu$b) for elastic photoproduction of 
 $\Upsilon(1s)$ in ultraperipheral PbPb collisions at the LHC.}
\begin{tabular}[c]{|lc|c| c c c}
\hline
PDF                 && Cross Section  \\
\hline
MSTW08              && 189  \\		      
EPS08               && 99  \\		       
EPS09	            && 130  \\		      
HKN07               && 146 \\
\hline	               
\end{tabular}
\label{tupsiPbPb}
\end{table}

Although not quite as dramatic as in the case of $J\Psi$ here also 
one can see the effect on the cross sections of the 
quadratic dependence on gluon distribution. The MSTW08 
(no-shadowing) cross section is reduced by approximately $48\%$ 
by the strong shadowing in EPS08, by about $31\%$ by the 
relatively moderate shadowing in EPS09, and by $23\%$ by the rather weak 
gluon shadowing present in HKN07. Thus a consideration of the cross  
section can serve as a good probe of gluon shadowing as well as 
an efficient discriminator of the different gluon shadowing scenarios
considered in the current study.

In order to further exhibit the constraining potentials of $\Upsilon$
photoproduction we display the rapidity distributions in  
Fig.~\ref{upsirapPbPb} for the four different gluon shadowing scenarios. 

\begin{figure}[!htb]
\begin{center} 
\includegraphics[width=8.5cm, height=8.5cm, angle=270]{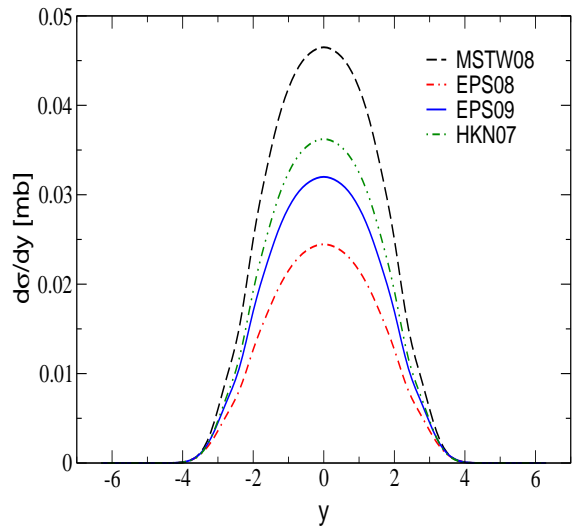}
\end{center}
\caption[...]{(Color online) Total rapidity distributions of exclusive
  photoproduction of $\Upsilon(1s)$ in PbPb collisions at the LHC in the
  modified hard sphere density distribution approximation. Dashed
  line depicts result using the MSTW08 gluon distribution (no nuclear
  modifications). Solid, dash-dotted, and dash-double-dotted lines are results
  from nuclear-modified gluon distributions from EPS09, EPS08, and
  HKN07 parton distributions respectively.}
\label{upsirapPbPb}
\end{figure}
Shadowing remains the relevant nuclear modification for practically
the entire rapidity range shown in the figure, and is markedly
manifested in the interval $-2 < y < 2$. Thus rapidity distribution 
in this interval should be a good discriminator of gluon shadowing strength. 

\subsection{Theoretical errors and uncertainties in parton distributions}
 From a consideration of the general structure of the quantities 
(cross sections and rapidity distributions) considered in this study, 
three major sources of theoretical errors can be readily identified: 
\begin{itemize}
\item accuracy of the relevant expressions for the photon flux, 
\item higher-order corrections (neglect of terms greater than 
leading order in the case of photoproduction of heavy quarks, the 
accuracy of mopping up higher-order corrections with data from HERA 
in the case of production of $J/\Psi$ and $\Upsilon$), 
\item uncertainties in the parton distributions. 
\end{itemize} 
Here we give a brief qualitative comment on the third, i.e. the
effects of uncertainties in parton distributions on our calculations. 
Let us first consider uncertainties in nuclear parton distributions (nPDs).
This relevant for the dominant direct contribution in photoproduction 
of heavy quarks and elastic production of vector mesons. 
Both HKN07 \cite{Hirai:2007sx} and EPS09 \cite{Eskola:2009uj} have 
facilities for estimating error in the determination of a 
nPDs-dependent quantity like the cross section 
due to the uncertainties in the nuclear parton distributions.
The HKN07 set achieves this through the availability of a set of
gradient functions while EPS09 has thirty error sets to be used in 
conjunction with free proton parton distributions error sets. 

For technical reasons we have not computed errors on cross sections 
using either of these two sets. Since this is a continuation of our 
earlier study, we have maintained the usage of the MSTW08 set for the 
free proton parton distributions while the EPS09 error evaluation 
is better handled with the CTEQ sets \cite{Pumplin:2002vw} of nucleon parton 
distributions. In the case of HKN07 we have 
used the nuclear modifications in our calculations whereas the
gradient terms are for the parton distributions proper. We thus 
advocate that the spread in the results involving the nuclear 
parton distributions be taken as loosely indicative of the effect of 
the uncertainties in the nuclear PDs. 

We now consider the resolved component of photoproduction of heavy 
quarks where photon parton distributions are required. Among the
available sets of photon parton distributions, only the CJK set has 
the facility to compute errors from uncertainties in photon PDs in
terms of a tolerance parameter $T$ \cite{Cornet:2004ng,Jankowski:2003mt}, 
with the CJK2 uncertainty bands generated with $T=5$.  
Thus, in principle, one needs only the CJK2 best 
fit and the associated error sets in order to compute the errors in
the resolved contribution due to uncertainties in $\gamma$PDs. The 
problem here is that the SaS1d distributions seems to lie consistently 
outside the CJK2 parton distributions error bands (further details can
be found in \cite{Jankowski:2003mt}). 
A reflection of this trend can be seen in our 
results for the resolved component where the CJK2 values are
consistently about a factor of two greater than the SaS1d values.
As a test we have used the CJK2 error sets in conjunction 
with EPS09 to evaluate the uncertainty on the $c\bar{c}$ resolved
cross section in PbPb collisions, with the result 
$\sigma^{res}_{CJK2} = 212.6 \pm 10.7\,T$ mb. The corresponding cross
sections using SaS1d and GRV are $\sigma^{res}_{SaS1d} = 101.1$ mb and 
$\sigma^{res}_{GRV} = 169.3$ mb respectively.   
Thus the GRV result is already within the error limits of the CJK for 
$T \simeq 4$ whereas one needs $T \simeq 10.4$ in order to accommodate
the SaS1d result. In view of this situation and since we have no
cogent reason to exclude the SaS1d distributions, we have presented results
from all three sets of $\gamma$PDs and adopt the viewpoint espoused 
above that the spread in results should be taken rather loosely as an 
indicator of the effects of the uncertainties in the  $\gamma$PDs. 
\section{Conclusions}
\label{conc}
In the present study we have considered photoproduction of
heavy quarks ($c\bar{c}$ and $b\bar{b}$) and elastic photoproduction 
of vector mesons ($J/\Psi$ and $\Upsilon(1s)$) in ultraperipheral 
proton-lead (pPb) and lead-lead (PbPb) collisions at LHC.
Both the dominant direct component in photoproduction of heavy 
quarks and elastic production of heavy mesons are dependent on nuclear gluon
distributions, and could therefore be potentially useful in constraining
modifications such as shadowing and antishadowing in nuclear 
gluon distributions. The resolved component in photoproduction of 
heavy quarks depends on the distributions of light quarks and 
gluons in both photons and nuclei. Different sets of both nuclear and 
photon parton distributions with different attributes have been utilized.

In photoproduction of heavy quarks the parton distribution dependence is
linear and different modifications are superimposed due to the
integration over the momentum fraction $x$. Despite these limitations, 
both cross sections and rapidity distributions for $c\bar{c}$ in PbPb
collisions manifest appreciable sensitivity to
shadowing around midrapidity and a slight sensitivity to antishadowing 
at more forward and backward rapidities. Thus $c\bar{c}$
photoproduction offers good constraining potential for shadowing, and
a somewhat less potential for antishadowing. Although photoproduction of 
$b\bar{b}$ is less sensitive to modifications than $c\bar{c}$, the 
influence of shadowing is evident around midrapidity, and it thus
offers some constraining ability for shadowing. While both $c\bar{c}$ and 
$b\bar{b}$ total photoproduction cross sections and rapidity
distributions in pPb collisions show little sensitivity
to nuclear modifications, the rapidity asymmetry ratios in some select 
intervals do exhibit significant sensitivity. The resolved components are 
appreciable, especially for $b\bar{b}$ and are heavily dependent on
the choice of $\gamma$PDs. Thus it seems feasible that they could be
of some use in constraining photon parton distributions.  

The outlook for constraining gluon shadowing is even better in 
the case of vector meson production.   
Here the quadratic dependence on gluon modifications makes elastic
photoproduction of vector mesons particularly attractive for 
constraining purposes. The cross sections and rapidity 
distributions for both $J/\Psi$ and $\Upsilon(1s)$ photoproduction 
in PbPb collisions exhibit very good sensitivity to gluon shadowing. 
Thus both offer remarkable potential in constraining the shadowing 
component of nuclear gluon distributions. This is also true for 
$\Upsilon$ production and to a lesser extent $J\Psi$ production 
in pPb collisions.

\bigskip

We acknowledge support by the US Department of Energy grant  DE-FG02-08ER41533  and the Research Corporation.
%


\begin{thebibliography}{99}
%
\bibitem{Jackson}
E. Fermi, Z. Physik  {\bf 29}, 315 (1924); Nuovo Cimento {\bf 2}, 143 (1925).

\bibitem{Bertulani:1988}
C.~A. Bertulani and G.~Baur, {Phys. Rep.} {\bf 163}, 299 (1988).

\bibitem{Cahn:1990jk}
R.~N. Cahn and J. D. Jackson, {Phys.\ Rev.} D {\bf 42}, 3690 (1990).

\bibitem{Baur:1990fx}
G. Baur and L. G. Ferreira Filho,
{Nucl.\ Phys.} A {\bf 518}, 786 (1990).

\bibitem {KN99}S. Klein and J. Nystrand, {Phys. Rev. C } {\bf 60}, 014903
(1999).

\bibitem{Bertulani:1999cq}
  C.~A.~Bertulani and D.~S.~Dolci,
  Nucl.\ Phys.\  A {\bf 674}, 527 (2000).
  
  
\bibitem{Goncalves:2001vs}
  V.~P.~Goncalves and C.~A.~Bertulani,
  Phys.\ Rev.\  C {\bf 65}, 054905 (2002).

\bibitem{KNV02}
S.~R.~Klein, J.~Nystrand and R.~Vogt,
Phys.\ Rev.\ C {\bf 66}, 044906 (2002).

\bibitem{Goncalves:2003is}
  V.~P.~Goncalves and M.~V.~T.~Machado,
  Eur.\ Phys.\ J.\  C {\bf 31}, 371 (2003).

\bibitem{Bertulani:2005ru}
  C.~A.~Bertulani, S.~R.~Klein and J.~Nystrand,
  Ann.\ Rev.\ Nucl.\ Part.\ Sci.\  {\bf 55}, 271 (2005).

\bibitem{Baltz:2007kq}
  A.~J.~Baltz {\it et al.},
  Phys.\ Rept.\  {\bf 458}, 1 (2008).

\bibitem{AyalaFilho:2008zr}
  A.~L.~Ayala Filho, V.~P.~Goncalves and M.~T.~Griep,
  Phys.\ Rev.\  C {\bf 78}, 044904 (2008).

\bibitem{Salgado:2011wc}
  C.~A.~Salgado, Ed., ``Proton-Nucleus Collisions at the LHC: Scientific Opportunities and Requirements", arXiv:1105.3919.

\bibitem{Adeluyi:2011rt} 
  A.~Adeluyi and C.~Bertulani,
  Phys.\ Rev.\ C {\bf 84}, 024916 (2011)


\bibitem{Drees:1988pp} 
  M.~Drees and D.~Zeppenfeld,
  Phys.\ Rev.\ D {\bf 39}, 2536 (1989).


\bibitem{Gluck:1978bf}
  M.~Gluck and E.~Reya,
  Phys.\ Lett.\  B {\bf 79}, 453 (1978).

\bibitem {JonesWyld}\textrm{L. M. Jones and H. W. Wyld, Phys. Rev. D } {\bf 17}, 759
(1978).

\bibitem {FriStreng78}\textrm{H. Fritzsch and K. H. Streng, Phys. Lett. B}
{\bf 72}, 385 (1978).

\bibitem{Gluck:1977zm} 
  M.~Gluck, J.~F.~Owens and E.~Reya,
  Phys.\ Rev.\ D {\bf 17}, 2324 (1978).

\bibitem{Combridge:1978kx} 
  B.~L.~Combridge,
  Nucl.\ Phys.\ B {\bf 151}, 429 (1979).

\bibitem{Brock:1993sz} 
  R.~Brock {\it et al.}  [CTEQ Collaboration],
  Rev.\ Mod.\ Phys.\  {\bf 67}, 157 (1995).

\bibitem{strikman_plb}
L.~Frankfurt, M.~Strikman and M.~Zhalov,
Phys.\ Lett.\ B {\bf 540}, 220 (2002).

 \bibitem{per4}
  V.~P.~Goncalves and M.~V.~T.~Machado,
  Eur.\ Phys.\ J.\  C {\bf 40}, 519 (2005).

\bibitem{vicmag_prd2008}
  V.~P.~Goncalves and M.~V.~T.~Machado,
  Phys.\ Rev.\  D {\bf 77}, 014037 (2008).

\bibitem{ivanov_kop}
  Yu.~P.~Ivanov, B.~Z.~Kopeliovich and I.~Schmidt,
  arXiv:0706.1532 [hep-ph].

\bibitem{strikman_jhep}
  L.~Frankfurt, V.~Guzey, M.~Strikman and M.~Zhalov,
  JHEP {\bf 0308}, 043 (2003).

\bibitem{klein_prl}
S.~R.~Klein and J.~Nystrand,
Phys.  Rev. Lett.  {\bf 92}, 142003 (2004).

\bibitem{vicmag_prc}
  V.~P.~Goncalves and M.~V.~T.~Machado,
  Phys. Rev. C {\bf 73}, 044902 (2006).

\bibitem {Ryskin}\textrm{M. G. Ryskin, Z. Phys., C } {\bf 57}, 89
  (1993).

\bibitem{Brodsky:1994kf}
  S.~J.~Brodsky, L.~Frankfurt, J.~F.~Gunion, A.~H.~Mueller and M.~Strikman,
  Phys.\ Rev.\  D {\bf 50}, 3134 (1994)

\bibitem{Ryskin:1995hz}
  M.~G.~Ryskin, R.~G.~Roberts, A.~D.~Martin and E.~M.~Levin,
  Z.\ Phys.\  C {\bf 76}, 231 (1997)

\bibitem{Frankfurt:1997fj}
  L.~Frankfurt, W.~Koepf and M.~Strikman,
  Phys.\ Rev.\  D {\bf 57}, 512 (1998)

\bibitem{Adloff:2000vm}
  C.~Adloff {\it et al.}  [H1 Collaboration],
  Phys.\ Lett.\  B {\bf 483}, 23 (2000)

\bibitem{Breitweg:1998ki}
  J.~Breitweg {\it et al.}  [ZEUS Collaboration],
  Phys.\ Lett.\  B {\bf 437}, 432 (1998)

\bibitem{Chekanov:2009zz}
  S.~Chekanov {\it et al.}  [ZEUS Collaboration],
  Phys.\ Lett.\  B {\bf 680}, 4 (2009)

\bibitem{DeJager:1974dg}
  C.~W.~De Jager, H.~De Vries and C.~De Vries,
  Atom.\ Data Nucl.\ Data Tabl.\  {\bf 14}, 479 (1974).

\bibitem{DN} K. T. R. Davies and J. R. Nix, Phys. Rev. 
{\bf C14}, 1977 (1976).

\bibitem{Aubert:1983xm}
  J.~J.~Aubert {\it et al.}  [European Muon Collaboration],
  Phys.\ Lett.\  B {\bf 123}, 275 (1983).
%
\bibitem{Geesaman:1995yd}
  D.~F.~Geesaman, K.~Saito and A.~W.~Thomas,
  Ann.\ Rev.\ Nucl.\ Part.\ Sci.\  {\bf 45}, 337 (1995).
%
\bibitem{Piller:1999wx}
  G.~Piller and W.~Weise,
  Phys.\ Rept.\  {\bf 330}, 1 (2000).
%
\bibitem{Armesto:2006ph}
  N.~Armesto,
  J.\ Phys.\ G {\bf 32}, R367 (2006).

\bibitem{Kolhinen:2005az}
  V.~J.~Kolhinen,
 arXiv:hep-ph/0506287.

\bibitem{Eskola:1998df}
  K.~J.~Eskola, V.~J.~Kolhinen and C.~A.~Salgado,
  Eur.\ Phys.\ J.\ C {\bf 9}, 61 (1999).
%
\bibitem{deFlorian:2003qf}
  D.~de Florian and R.~Sassot,
  Phys.\ Rev.\  D {\bf 69}, 074028 (2004).
%
\bibitem{Shad_HKN}
  M.~Hirai, S.~Kumano and T.~H.~Nagai,
  Phys.\ Rev.\  C {\bf 70}, 044905 (2004);
  Nucl.\ Phys.\ Proc.\ Suppl.\  {\bf 139}, 21 (2005).
%
\bibitem{Hirai:2007sx}
  M.~Hirai, S.~Kumano and T.~H.~Nagai,
  Phys.\ Rev.\  C {\bf 76}, 065207 (2007).

\bibitem{Eskola:2008ca}
  K.~J.~Eskola, H.~Paukkunen and C.~A.~Salgado,
  JHEP {\bf 0807}, 102 (2008).
%
\bibitem{Eskola:2009uj}
  K.~J.~Eskola, H.~Paukkunen and C.~A.~Salgado,
  JHEP {\bf 0904}, 065 (2009).

\bibitem{Schienbein:2009kk} 
  I.~Schienbein, J.~Y.~Yu, K.~Kovarik, C.~Keppel, J.~G.~Morfin, F.~Olness and J.~F.~Owens,
  Phys.\ Rev.\ D {\bf 80}, 094004 (2009)

\bibitem{Stavreva:2010mw} 
  T.~Stavreva, I.~Schienbein, F.~Arleo, K.~Kovarik, F.~Olness, J.~Y.~Yu and J.~F.~Owens,
  JHEP {\bf 1101}, 152 (2011)

\bibitem{Kovarik:2010uv} 
  K.~Kovarik, I.~Schienbein, F.~I.~Olness, J.~Y.~Yu, C.~Keppel, 
J.~G.~Morfin, J.~F.~Owens and T.~Stavreva,
  Phys.\ Rev.\ Lett.\  {\bf 106}, 122301 (2011)

\bibitem{Frankfurt:2003zd}
  L.~Frankfurt, V.~Guzey and M.~Strikman,
  Phys.\ Rev.\ D {\bf 71}, 054001 (2005).

\bibitem{Frankfurt:2011cs} 
  L.~Frankfurt, V.~Guzey and M.~Strikman,
  arXiv:1106.2091 [hep-ph].


\bibitem{Martin:2009iq}
  A.~D.~Martin, W.~J.~Stirling, R.~S.~Thorne and G.~Watt,
  Eur.\ Phys.\ J.\  C {\bf 63}, 189 (2009).

\bibitem{Duke:1982bj} 
  D.~W.~Duke and J.~F.~Owens,
  Phys.\ Rev.\ D {\bf 26}, 1600 (1982).

\bibitem{Drees:1984cx} 
  M.~Drees and K.~Grassie,
  Z.\ Phys.\ C {\bf 28}, 451 (1985).

\bibitem{Abramowicz:1991yb} 
  H.~Abramowicz, K.~Charchula and A.~Levy,
  Phys.\ Lett.\ B {\bf 269}, 458 (1991).

\bibitem{Hagiwara:1994ag} 
  K.~Hagiwara, M.~Tanaka, I.~Watanabe and T.~Izubuchi,
  Phys.\ Rev.\ D {\bf 51}, 3197 (1995)

\bibitem{Gluck:1991ee} 
  M.~Gluck, E.~Reya and A.~Vogt,
  Phys.\ Rev.\ D {\bf 45}, 3986 (1992).

\bibitem{Gluck:1994tv} 
  M.~Gluck, E.~Reya and M.~Stratmann,
  Phys.\ Rev.\ D {\bf 51}, 3220 (1995).

\bibitem{Gluck:1991jc} 
  M.~Gluck, E.~Reya and A.~Vogt,
  Phys.\ Rev.\ D {\bf 46}, 1973 (1992).

\bibitem{Gordon:1996pm} 
  L.~E.~Gordon and J.~K.~Storrow,
  Nucl.\ Phys.\ B {\bf 489}, 405 (1997)
  [hep-ph/9607370].

\bibitem{Gordon:1991tk} 
  L.~E.~Gordon and J.~K.~Storrow,
  Z.\ Phys.\ C {\bf 56}, 307 (1992).

\bibitem{Schuler:1996fc} 
  G.~A.~Schuler and T.~Sjostrand,
  Phys.\ Lett.\ B {\bf 376}, 193 (1996)
  [hep-ph/9601282].

\bibitem{Schuler:1995fk} 
  G.~A.~Schuler and T.~Sjostrand,
  Z.\ Phys.\ C {\bf 68}, 607 (1995)

\bibitem{Aurenche:1994in} 
  P.~Aurenche, J.~P.~Guillet and M.~Fontannaz,
  Z.\ Phys.\ C {\bf 64}, 621 (1994)

\bibitem{Aurenche:1992sb} 
  P.~Aurenche, P.~Chiappetta, M.~Fontannaz, J.~P.~Guillet and E.~Pilon,
  Z.\ Phys.\ C {\bf 56}, 589 (1992).

\bibitem{Cornet:2002iy} 
  F.~Cornet, P.~Jankowski, M.~Krawczyk and A.~Lorca,
  Phys.\ Rev.\ D {\bf 68}, 014010 (2003)

\bibitem{Cornet:2004ng} 
  F.~Cornet, P.~Jankowski and M.~Krawczyk,
  Acta Phys.\ Polon.\ B {\bf 35}, 2215 (2004)
  [hep-ph/0404244].

\bibitem{Cornet:2004nb} 
  F.~Cornet, P.~Jankowski and M.~Krawczyk,
  Phys.\ Rev.\ D {\bf 70}, 093004 (2004)

\bibitem{Strikman:2005yv}
  M.~Strikman, R.~Vogt and S.~N.~White,
  Phys.\ Rev.\ Lett.\  {\bf 96}, 082001 (2006)
  [arXiv:hep-ph/0508296].

\bibitem{Pumplin:2002vw} 
  J.~Pumplin, D.~R.~Stump, J.~Huston, H.~L.~Lai, P.~M.~Nadolsky and W.~K.~Tung,
  JHEP {\bf 0207}, 012 (2002)

\bibitem{Jankowski:2003mt} 
  P.~Jankowski,
  JHEP {\bf 0405}, 055 (2004)
%
\end{thebibliography}
\end{document}